# Effects of the choice of the exchange-correlation functional on high-temperature superconductivity predictions in covalent-cage $M(BN)_5$ family

Zhongyu Wan[1]*, Guo-Hua Zhong[2][3], Cheng-wei Ju[4], Ruiqin Zhang[1][5]**

and Hai-Qing Lin[6]***

[1] Department of Physics, City University of Hong Kong, Hong Kong SAR 999077, People's Republic of China

[2] Shenzhen Institute of Advanced Technology, Chinese Academy of Sciences, Shenzhen 518055, People's Republic of China

[3] University of Chinese Academy of Sciences, Beijing 100049, People's Republic of China

[4] Pritzker School of Molecular Engineering, The University of Chicago, Chicago, Illinois 60637, USA

[5] Advanced Energy Storage Technology Research Center, Shenzhen Polytechnic University, Shenzhen 518055, People's Republic of China

[6] School of Physics, Zhejiang University, Hangzhou 310058, People's Republic of China

*Email: zywan5-c@my.cityu.edu.hk
**Email: aprqz@cityu.edu.hk
***Email: hqlin@zju.edu.cn

## Abstract

The recent experimental synthesis of covalent cage superconductors has ignited hopes for ambient-pressure high-temperature superconductivity, while the computational strategy for discovering new materials poses challenges for accurately estimating superconducting properties. Here, we quantified the response of electron-phonon coupling (EPC) properties in $M(BN)_5$ superconductors to different exchange-correlation (XC) effects using high-throughput first-principles calculations. We found that the metallic covalent bonds responsible for superconductivity exhibit anomalously high sensitivity to functional selection uncertainty, distinguishing them from conventional superconductors mediated by metallic bonding. The widely used PBE functional systematically overestimates lattice volumes, inducing phonon softening, enhanced EPC, and consequent Peierls instability. In contrast, PBEsol, which is more accurate through enhanced treatment of short-range interactions, eliminates imaginary frequencies in $Rb(BN)_5$ that appear at the PBE level, yielding an outstanding superconducting critical temperature ($T_c$) of 113 K at ambient pressure and 138 K at 2 GPa. Enhanced electron localization and short-range correlation in covalent bonds manifest through dual mechanisms: phonon hardening at the lattice scale and band widening at the electronic scale, which collectively suppress Peierls distortion and delocalize antibonding states. Pressure-dependent analysis reveals that compression attenuates phonon discrepancies between XC functionals, whereas electronic properties remain notably sensitive to XC-induced perturbations of fine electronic bands near the Fermi level. This work demonstrates the functional-dependent sensitivity of the predicted superconducting search space in density-functional predictions, highlights how subtle XC functional choices can shift strongly coupled frameworks across dynamical stability boundaries, and provides critical benchmarks and references for future computational superconductor design.

## 1. **Introduction**

The resistance-free behavior of mercury below 4.2 K, observed by Onnes in 1911, marked a milestone in physics as the discovery of the first superconductor.[1] Subsequent decades of investigation into its microscopic mechanism led to the Bardeen-Cooper-Schrieffer theory,[2] establishing the phonon-mediated pairing picture that explained conventional superconductivity. By the 1990s, advances in density functional theory (DFT) and high-performance computing enabled quantitative characterization of electron-phonon coupling through perturbation theory,[3] opening avenues for computationally guided superconductor discovery—particularly the pursuit of critical temperatures exceeding the liquid-nitrogen threshold, with transformative implications for energy transmission and quantum computing.[4,5]

Early first-principles results were promising.[6] Building on Ashcroft's insight into "metallic hydrogen,"[7] DFT successfully predicted numerous high-$T_c$ hydride superconductors subsequently confirmed experimentally,[8] sparking the recent "*hydride rush*."[9] Classic examples include $H_3S$ and $LaH_{10}$ with near-room-temperature $T_c$ under pressure.[10,11] Yet significant discrepancies between experimental measurements and theoretical predictions in these hydrides—later reconciled through quantum anharmonic corrections for proton dynamics—highlight accurate $T_c$ prediction as an outstanding challenge in *ab initio* superconductor design.[12,13] For non-hydride BCS superconductors, while some theoretical results agree well with experiment, exceptions persist even with advanced Migdal-Eliashberg methods and dynamic Coulomb pseudopotentials.[14] Emerging evidence attributes these $T_c$ estimation variations to XC effects—a fundamental yet often overlooked consideration in

DFT workflows,[15] historically selected arbitrarily.[16]

The treatment of electron density gradients ($\nabla\rho$) underlies functional applicability,[17,18] particularly as BCS Cooper pairs form from metalized chemical bonds at the Fermi level ($E_f$). Although PBE yields $T_c$ closer to experiment for specific systems such as Al and Nb, PBEsol achieves smaller mean absolute errors across broader test sets—with advantages particularly pronounced in solids dominated by short-range interactions.[15,19,20] The recent experimental synthesis of $SrB_3C_3$ has highlighted the potential of covalent-cage compounds for ambient-pressure high-temperature superconductivity,[21,22] motivating extensive PBE-based predictions of $M(BN)_5$,[23] $M(BN)_3$,[24] $MC_6$,[25] $MC_{10}$,[26] $M(BC)_3$,[27-29] and $MB_6$ (M is various elements).[30] Yet benchmark tests reveal PBEsol's superior performance under pressure, with XC discrepancies amplifying at lower pressures.[31] We therefore hypothesize that existing PBE calculations systematically overestimate covalent-cage bonding, leading to underestimated interatomic force constants, error-induced phonon softening,[32,33] and compromised EPC and phonon spectrum estimates—potentially shifting the predicted dynamic stability boundaries and critical temperature estimates in strongly coupled covalent frameworks.

This study investigates ambient-pressure XC effects on covalent BCS superconductor predictions. Specifically, we address: (i) the systematic variations in EPC strength ($\lambda$), logarithmic average phonon frequency ($\omega_{log}$), and $T_c$ across common XC functionals; (ii) whether XC selection affects materials search space, masking additional high-$T_c$ superconductors; and (iii) the physical mechanisms underlying this exceptional sensitivity. This work underscores the importance of adaptive functional selection in computational workflows and provides critical methodological guidance for future studies.

## 2. **Results and Discussion**

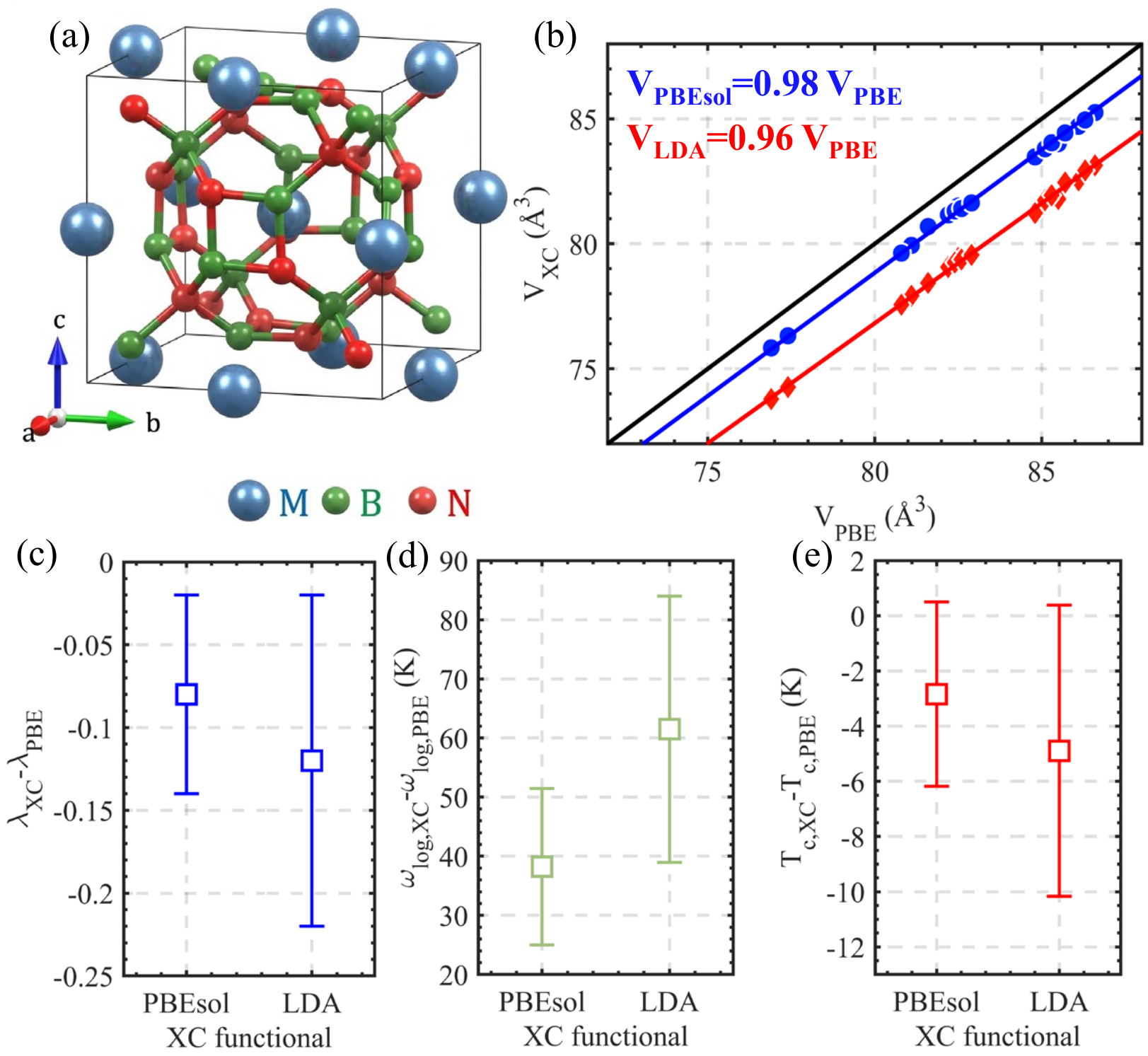

**Figure 1. (a) Crystal structure of $F\bar{4}3m$ symmetric M(BN)$_5$. (b) Linear fit of PBE versus PBEsol and LDA volumes for M(BN)$_5$. Error bars for (c) $\lambda$, (d) $\omega_{log}$, and (e) $T_c$ deviations from PBE values across different XC functionals.**

The functional hierarchy LDA → GGA → Meta-GGA → hybrid functionals reflects increasing accuracy,[34] yet strong nonlinear mathematical forms incompatible with linear-response EPC calculations render Meta-GGAs and hybrids impractical.[35,36] Although combining frozen-phonon PBE0 force constants with perturbation-theory PBE-based EPC matrix elements has been attempted for $H_3S$, mismatched nonlinear XC responses between methods exacerbated $T_c$ errors relative to experiment.[37] Accessibility and reliability restrict current *ab initio* superconductivity methods to LDA, GGA-PBE, and solid-optimized GGA-PBEsol,[15] which we adopt as our standard. The B-N covalent network in M(BN)$_5$ (Figure 1(a)) provides an ideal platform for comparing these functionals, with covalent-bond sensitivity to XC effects warranting re-evaluation of existing predictions, given that metalized covalent bonds constitute the superconducting origin.[38,39]

High-throughput screening across the periodic table (excluding transition metals, lanthanides, and actinides to eliminate strong electronic correlation and spin-orbit coupling effects)[40,41] reveals that LDA and PBEsol yield lattice dynamical stability estimates consistent with reported PBE results for M = Na, Al, Ga, In, Tl, Sn, Pb, As, O, Sb, Bi, S, Se, Te, F, Cl, Br, and I.[42] Compared to PBE, PBEsol underestimates volumes by 2% ($\Delta V_{PBEsol}$ = 1.21 ± 0.15 Å$^3$/cell), while LDA expands this underestimation to 4% (3.30 ± 0.18 Å$^3$/cell) (Figure 1(b)). For EPC properties, PBEsol underestimates $\lambda$ by 0.08 ± 0.06 relative to PBE, with LDA showing larger underestimation of 0.12 ± 0.10 (Figure 1(c)). For logarithmic average phonon frequencies, PBEsol and LDA overestimate $\omega_{log}$ by 38.24 ± 13.23 K and 61.48 ± 22.51 K (Figure 1(d)), respectively. Interatomic force constants are intrinsically tied to equilibrium positions,[43] with XC effects inducing divergent relaxed lattice configurations; consequently, the $\omega_{log}$ overestimation by PBEsol and LDA likely stems from their systematic volume underestimation. A wealth of existing benchmarks consistently reveal the relationship $V_{LDA}$ < $V_{PBEsol}$ < $V_{PBE}$.[44,45] This stems from the LDA relies solely on the local electron density and overlooks the $\nabla\rho$, leading to an inadequate description of electron-electron correlation.[46] Such a simplification underestimates the electron-electron repulsion in real systems, consequently resulting in smaller lattice. In contrast, lattice constants with PBE tend to be overestimated compared to those from PBEsol. As for its exchange enhancement factor:

$$F_X(s) = 1 + \kappa - \kappa \big/ \left(1 + \mu s^2 / \kappa\right) \tag{1}$$

Where PBEsol reduced $\mu$=0.123 < 0.219 of PBE,[47] diminishing exchange energy response to $\nabla\rho$.[48]

For correlation energy:

$$H\left(\rho, s\right) = \gamma\phi^3 \ln\left(1 + \beta t^2 \big/ \gamma\left(1 + \beta t^2 / \gamma^2\right)\right) \tag{2}$$

Notably, PBEsol features a smaller $\beta$ value of 0.046 compared to PBE's 0.067,[49] which fine-tunes the electron correlation effects in solids and makes it well-suited for solid-state chemical environments.[50] Notably, provided that the exchange-correlation form remains identical, different pseudopotentials are expected to yield virtually identical physical descriptions, as their primary distinction lies in the treatment of core electrons.[51] Although subtle discrepancies among pseudopotentials might arise under extreme pressure due to core overlaps, core-electron variations under ambient pressure are far too small to affect physical predictions.[52] Indeed, standard benchmarks have established that different pseudopotential databases exhibit negligible differences for a given system,[53] where any noticeable deviation is typically attributable to insufficient kinetic energy cutoffs. In sharp contrast, variations stemming from the choice of XC functional are substantially more pronounced. To explicitly verify this, we compared the performance of the PAW pseudopotential from PSlibrary against the ONCV pseudopotential from SG15 for $Rb(BN)_5$ (Figure S4). The test confirms that pseudopotential choice does not alter our physical conclusions, demonstrating that the primary source of computational uncertainty lies in the exchange-correlation effects rather than pseudopotential artifacts.

Compared to widely used PBE, PBEsol's accurate treatment of interatomic distances—critical for bond strengths and forces—yields lattice properties most consistent with inelastic scattering measurements.[54,55] Given the central role of phonon-mediated mechanisms in BCS superconductivity,[56] PBEsol results carry greater predictive weight. To establish a structurally aligned baseline for describing covalent-cage superconductors, the recently synthesized ambient-pressure $CaB_3C_3$ was selected as a benchmark system.[57] Under identical computational parameters, the $T_c$ predicted by PBE is $6\ \mathrm{K}$, which deviates significantly from the experimental value of $4.6\ \mathrm{K}$ compared to the PBEsol result of $4.1\ \mathrm{K}$ (Figure S2 and Table S6). Furthermore, for the ambient-pressure cell volume, PBEsol exhibits a remarkably small deviation of less than $0.2\ \mathrm{Å}^3/\mathrm{cell}$, whereas the error for PBE reaches as high as $3.3\ \mathrm{Å}^3/\mathrm{cell}$. These benchmarking results clearly demonstrate that the PBEsol functional provides substantially higher accuracy and reliability for covalent-cage frameworks.

To quantify XC effects on covalent bond vibrational frequencies, we employ Badger's rule:[58]

$$k = \frac{C}{(r_e - d)^n} \tag{3}$$

where $k$ is the harmonic force constant, $r_e$ is the equilibrium bond length, $d$ is an empirical parameter specific to the atom pair, $C$ is an empirical constant, and $n$=3 in the standard rule (later refined to 5).[59] Combining with the harmonic oscillator model:[60]

$$\omega = \frac{1}{2\pi c}\sqrt{\frac{k}{\mu}} \Rightarrow k = 2\pi c\omega^2 \mu = \frac{C}{(r_e - d)^5} \tag{4}$$

where $c$ is the light speed and $\mu$ is the reduced mass. This yields:

$$r_e = \frac{C}{\left(2\pi c\omega^2 \mu\right)^{1/5}} + d \tag{5}$$

Taking the logarithm of above mathematical relation:

$$\ln r_e = \ln\frac{C}{\left(2\pi c\mu\right)^{1/5}} - \frac{2}{5}\ln\omega + \ln d \tag{6}$$

Further differentiation yields:

$$\frac{d\omega}{\omega} = -\frac{5}{2}\frac{dr_e}{r_e} \tag{7}$$

In fact, the variations induced by the XC effect are minute. Therefore, it is reasonable to assume that $d\omega \approx \Delta\omega$ and $dr_e \approx \Delta r_e$, allowing the differential forms to be rewritten as:

$$\frac{\Delta\omega}{\omega} = -\frac{5}{2}\frac{\Delta r_e}{r_e} \tag{8}$$

Therefore, the formula (8) can be employed to estimate the proportional change in phonon frequency. The fact that the PBEsol functional "compresses" the volume by 2% implies a 0.67% reduction in atomic spacing, resulting in $\Delta r_e/r_e$ = -0.67%. Consequently, the frequency change can be calculated as $\Delta\omega/\omega$ = 1.68%. Similarly, for the LDA functional, $\Delta\omega/\omega$ = 3.36%. Covalent bonds respond more sensitively to pressure than metallic or ionic bonds, justifying our neglect of non-covalent interactions in $M(BN)_5$. Using $c$-BN Raman mode (1054 $cm^{-1}$) as reference for similar $sp^3$ B-N hybridization gives $\Delta\omega_{PBEsol}$ = 17.71 $cm^{-1}$ (25.50 K) and $\Delta\omega_{LDA}$ = 35.41 $cm^{-1}$ (51.00 K), consistent with calculated above $\Delta\omega_{log}$ values (38.24 ± 13.23 K and 61.48 ± 22.51 K)—supporting our computational validity. Although XC effects harden phonons, but reduced $\lambda$ lowers estimated $T_c$ relative to PBE-level (Figure 1(e)). Notably, excessive $\lambda$ indicates lattice instability;[61] PBEsol's systematic $\lambda$ underestimation may restore stability to unstable $M(BN)_5$ at PBE level. Re-examination reveals $K(BN)_5$ and $Rb(BN)_5$ are dynamically stable with PBEsol yet unstable with PBE—explaining their omission from prior screening (Figure 2 (a) and (b)).[42] At PBE level, $K(BN)_5$ exhibits −77 $cm^{-1}$ imaginary frequencies along Γ-X and K-Γ-L with Debye frequencies $\omega_D$ = 1118 $cm^{-1}$; and $Rb(BN)_5$ shows $\omega_D$ = 1079 $cm^{-1}$ with imaginary modes reaching −313 $cm^{-1}$ across the Brillouin zone. The thermal dynamic stability of $Rb(BN)_5$ is further corroborated by molecular dynamics simulations (Figure S3).

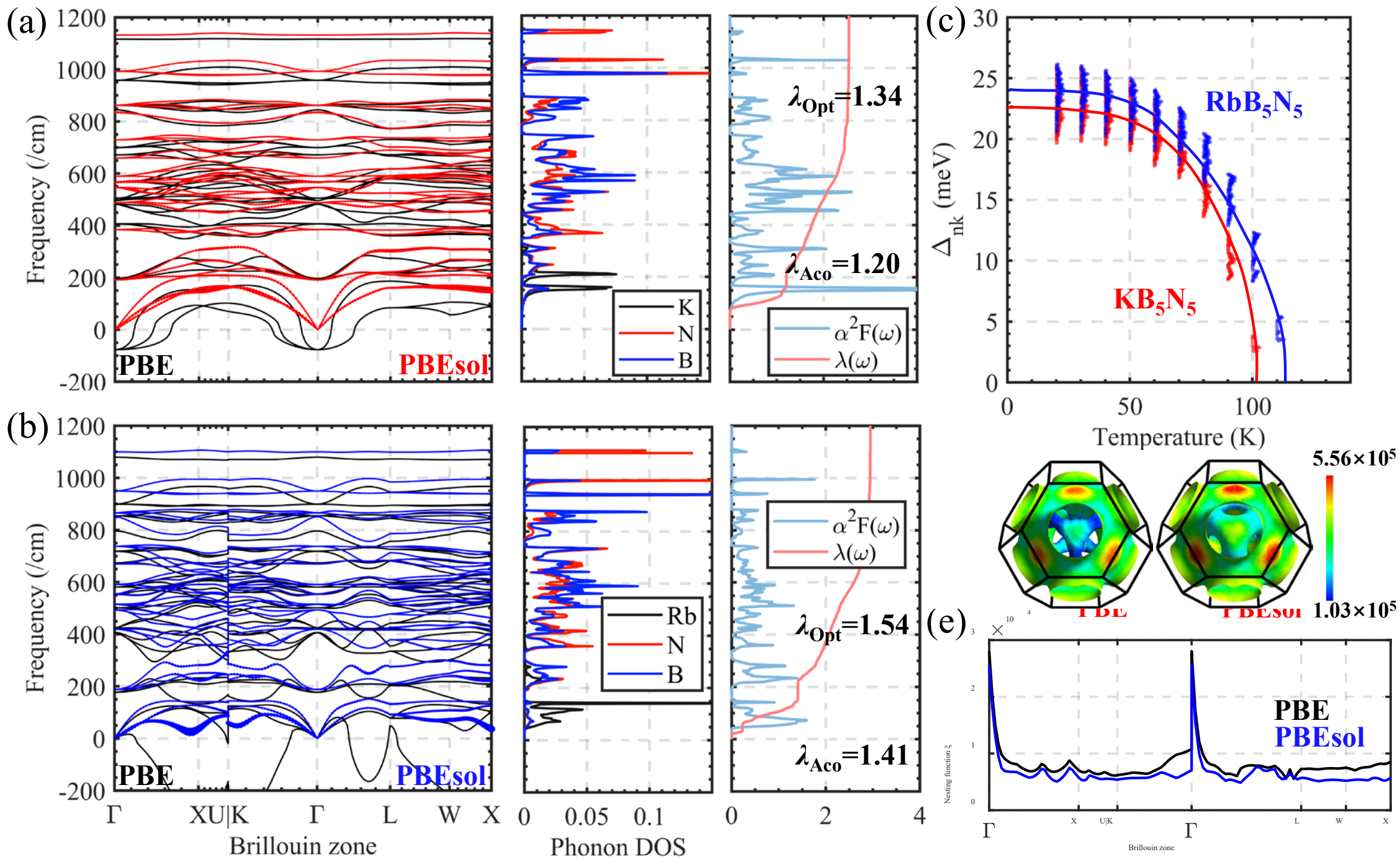

**Figure 2. Phonon dispersion curves, phonon density of states, and Eliashberg spectral functions for (a) $K(BN)_5$ and (b) $Rb(BN)_5$. (c)Full-bandwidth anisotropic superconducting gaps using a Coulomb pseudopotential $\mu^*$ = 0.1 and the PBEsol functional. (d) Fermi-velocity-weighted Fermi surfaces and (e) Fermi surface nesting functions of $Rb(BN)_5$ calculated at the PBE and PBEsol levels.**

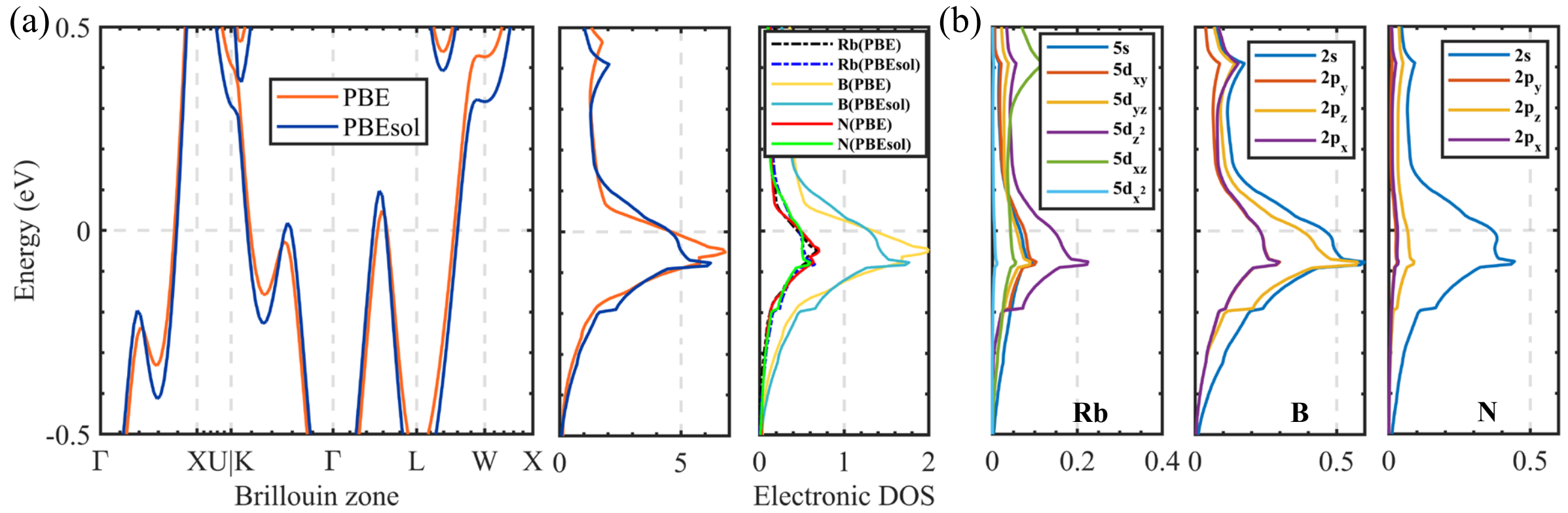

**Figure 3. (a) Electronic band structures，total density of states for Rb(BN)$_5$ at the PBE and PBEsol levels, respectively. (b) Atomic and angular momentum-resolved density of states for Rb(BN)$_5$ with PBEsol functional.**

Notably, PBEsol functional induces a blue shift in the $\omega_D$ of K(BN)$_5$ and Rb(BN)$_5$ by 31.67 K and 37.44 K, respectively, which aligns closely with the estimates derived from Badger's rule (equation (8)). Moreover, the imaginary frequency modes are eliminated by the XC effect, attributable to the stronger exchange interaction that generates a "stiffer" potential for covalent bond formation.[62] We also observe that for K(BN)$_5$, the acoustic branch contributes $\lambda_{Aco}$ = 1.20 (accounting for 47%), while the optical branch contributes $\lambda_{Opt}$ = 1.34 (53%) (Figure 2(a)). Similarly, $\lambda_{Aco}$ = 1.41 (48%) and $\lambda_{Opt}$ = 1.54 (52%) are found for Rb(BN)$_5$ (Figure 2(b)). Although the optical branches, comprising 30 phonon bands, dominate the electron-phonon coupling, while significant $\lambda_{Aco}$ from just three acoustic branches suggests that their contribution may be more crucial. Particularly, the Rb-dominated A' softening mode with 28 cm$^{-1}$ at the X point in Rb(BN)$_5$ is a potential factor contributing to its larger $\lambda$ compared to K(BN)$_5$. As a result, the superconducting gaps of K(BN)$_5$ and Rb(BN)$_5$ close at 102 K and 113 K, respectively (Figure 2(c)). In contrast, the best superconductor previously identified through PBE functional screening, F(BN)$_5$,[42] fails to surpass the liquid nitrogen threshold of 77 K, demonstrating XC effects indeed limit superconductor discovery. For the strongly coupled superconductor Rb(BN)$_5$, the Fermi velocity calculated at the PBE level ranges from $1.03 \times 10^5$ to $5.23 \times 10^5$ m/s, whereas PBEsol elevates this distribution to $1.35 \times 10^5$–$5.56 \times 10^5$ m/s (Figure 2(d)). Since Fermi velocity is defined as the energy gradient in reciprocal space,[63] this discrepancy indicates that PBE yields a weaker band dispersion and flatter bands compared to PBEsol. This behavior arises because the XC treatment in PBEsol naturally provides a tighter lattice constraint, inducing energy level splitting and degeneracy lifting, which consequently increases the Fermi velocity. Conversely, PBE introduces more electronic degeneracies near the Fermi level, triggering a band Jahn-Teller effect that drives lattice dynamical instability.[64] This XC-driven instability is further elucidated by Fermi surface nesting.[65] We observe that the nesting function $\xi(\mathbf{q})$ at the PBE level lies entirely above that of PBEsol, exhibiting 24 distinct nesting maximum compared to only 19 for PBEsol (Figure 2(e)). Therefore, the dynamical instability under PBE originates from a deepened Kohn anomaly caused by excessive electronic screening,[66] which severely weakens the interatomic restoring force constants and ultimately softens the phonon frequencies into imaginary modes. Furthermore, the greater number of nesting peaks under PBE indicates the presence of dense and diverse strong electron scattering channels in reciprocal space, implying a stronger driving force for Peierls distortions and exacerbating the tendency toward structural phase transitions.

The remarkably high $T_c$ achieved in $Rb(BN)_5$ underscores the necessity for deeper insights into its electronic structure. Analyses of the electronic band structure and the superconducting gap clearly reveal that only a single band crosses the Fermi level along the Γ-X and K-Γ-L-W paths (Figures 2(c) and 3(a)). Notably, the choice of XC functional significantly influences the trajectory of this band: PBEsol yields a steeper dispersion, while PBE produces a relatively flatter profile (Figure 3(a)). This indicates that PBEsol's enhanced description of short-range electron-electron interactions induces greater band broadening near $E_F$,[67] shifting the van Hove singularities to a deeper energy level of -0.08 eV and reducing their overall intensity. In contrast, the PBE functional generates a more pronounced DOS peak closer to the Fermi level (at -0.05 eV) due to heightened electron accumulation, which inherently amplifies the EPC strength. However, this excessive EPC induced by PBE triggers a Peierls instability,[68] manifesting as imaginary phonon modes. Conversely, the band splitting and enhanced dispersion captured by PBEsol successfully delocalize the electronic states, thereby maintaining lattice dynamical stability while preserving robust superconductivity.

It is crucial to clarify that enhanced EPC and proximity to structural instabilities are intrinsically linked physical characteristics in strongly coupled superconductors. The appearance of imaginary phonon modes under PBE and their resolution under PBEsol should not be interpreted as a simple physical breakdown of PBE. Rather, strongly coupled systems inherently operate near a lattice instability threshold. PBE's subtle overestimation of equilibrium cell volumes reduces interatomic force constants, thereby pushing the intrinsically soft phonon modes over the threshold into imaginary frequencies. Conversely, PBEsol provides a tighter lattice constraint through its improved description of short-range density gradients, preserving dynamic stability while retaining strong coupling. Therefore, these observations reflect the high susceptibility of strong-coupling predictions to functional-induced structural perturbations, rather than a straightforward methodological failure.

Atom-decomposed DOS reveals PBEsol most significantly reduces B DOS at -0.05 eV versus Rb and N atoms (Figure 3 (b)), with states originating from B-2s, B-2$p_z$, and N-2s orbitals. This indicates that the $\sigma$ covalent bond formed by B-sp hybrid orbitals and N-2s orbital is most significantly affected by the XC effect. Furthermore, crystal orbital Hamiltonian population analysis reveals that the $\sigma^*$ antibonding state within the N-centered [$NB_4$] covalent unit is the primary driving force for the metallize $E_f$ (Figure S1). Notably, the B-N bond length in [$NB_4$] unit calculated using PBEsol is 1.66 Å, which is shorter than the 1.68 Å under PBE. Therefore, the PBEsol functional intensifies the splitting of energy levels on both sides of $E_f$ by shortening the B-N bond and deepening the overlap of electron clouds. This reduction in the accumulated $\sigma^*$ antibonding states weakens the Peierls mechanism, thereby achieving stabilization.

Additionally, the central N atom in the [$NB_4$] unit, after providing its valence electrons to the four neighboring B atoms, retains one extra electron. In contrast, the central B atom in the [$BN_4$] unit, with only three valence electrons, exhibits electron deficiency. Consequently, the [$NB_4$] covalent unit contributes significantly more to the metallic character of the band structure than the [$BN_4$] unit (Figure S1). Furthermore, we investigated the isotropy of the superconducting gap in $Rb(BN)_5$ (Figure S6). Under the isotropic Eliashberg approximation, the superconducting gap closes at 107 K, which is only 6 K (~ 5%) lower than the anisotropic result (Figure 2(c)). This marginal deviation indicates that anisotropic electronic effects have a negligible impact on the estimation of $T_c$. In contrast, the McMillan–Allen–Dynes formula predicts a significantly lower $T_c$ of 49 K. This underestimation arises because the semi-empirical McMillan–Allen–Dynes equation is intrinsically unreliable for strong-coupling systems with $\lambda > 1.5$,[69] where numerical Eliashberg equations must be solved self-consistently.

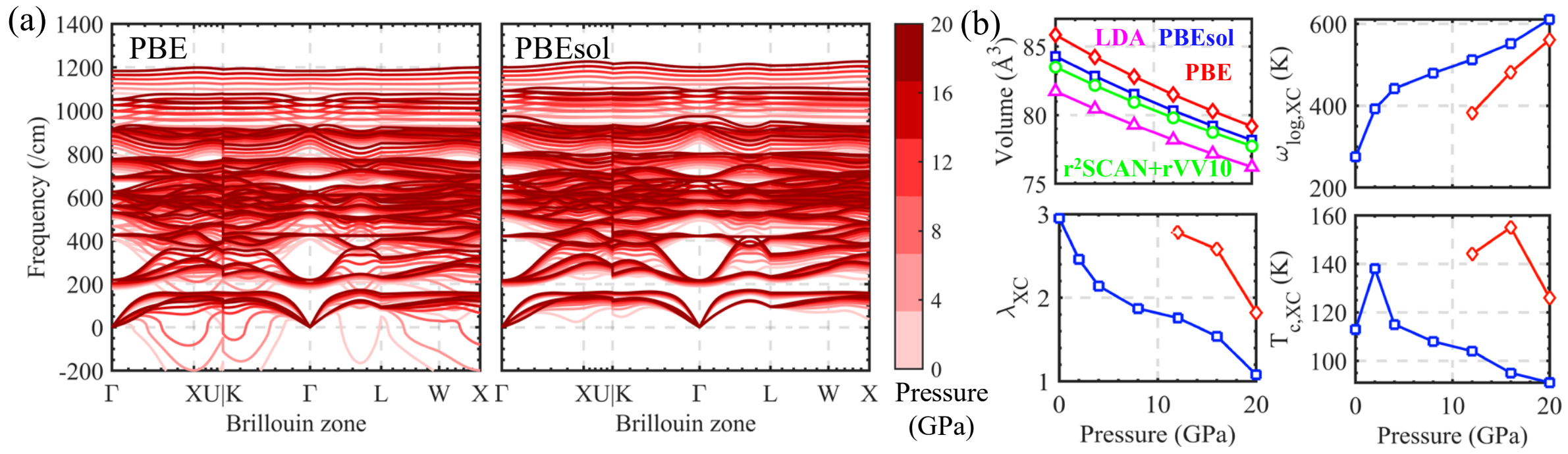

**Figure 4. (a) Phonon spectra of $Rb(BN)_5$ at 0–20 GPa using PBE and PBEsol functionals. (b) Volume, $\omega_{log}$, $\lambda$, and $T_c$ ($\mu^* = 0.1$) for compressed $Rb(BN)_5$ as functions of pressure.**

Pressure as an effective means to regulate the $T_c$, was applied within the range of 0 to 20 GPa (Figure 4(a)). Notably, the trends of $\lambda$ and $\omega_{log}$ with respect to pressure are consistent at both the PBE and PBEsol levels. Specifically, pressure enhances the interatomic force constants, hardens the phonons, reduces $\lambda$, and induces a blue shift in $\omega_{log}$ (Figure 4(b)).

However, a significant difference arises concerning the impact of XC effects on the dynamical stability of $Rb(BN)_5$. This disparity becomes evident within the lower pressure range of 0 to 12 GPa, while above 12 GPa, the behaviors tend to converge (Figure 4(a)). A similar reduction in discrepancy is observed in volume estimates, where the $\Delta V_{PBEsol}$ decreases from 1.6 Å³ (1.9%) at 0 GPa to 1.0 Å³ (1.3%) at 20 GPa (Figure 4(b)). Consequently, the divergence in $\omega_{log}$ estimates between the PBE and PBEsol also diminishes, with $\Delta\omega_{log,PBEsol}$ decreasing from 130.9 K (25.5%) at 12 GPa to 49.8 K (8.2%) at 20 GPa. This indicates that the pressure-induced reduction in the differences in $\omega_{log}$ between different XC functionals is consistent with the findings from XC tests on $H_3S$ at 200 GPa, which exhibited only a 7% error.[37] This convergence can be attributed to the pressure intensifies the overlap of electron clouds, causing different functionals to approach a similar uniform electron gas in the short range and reduces the dependence of interatomic force constants on the electron density gradient.[70]

In contrast, the disparity in $\lambda$ remains substantial for compressed $H_3S$ and $H_3Se$,[37,71,72] which can be attributed to the strong dependence of the electronic bands crossing the $E_f$ on the choice of XC functional. As demonstrated by the band structures of $Rb(BN)_5$ under ambient pressure and $H_3S$ under 200 GPa (Figure 3 (a)), the inherent differences among various functionals in terms of self-interaction and electron localization/delocalization are independent of pressure. This systematic bias causes fluctuations in the electronic bands within the Fermi window, thereby resulting in an exceptionally sensitive response of $\lambda$.

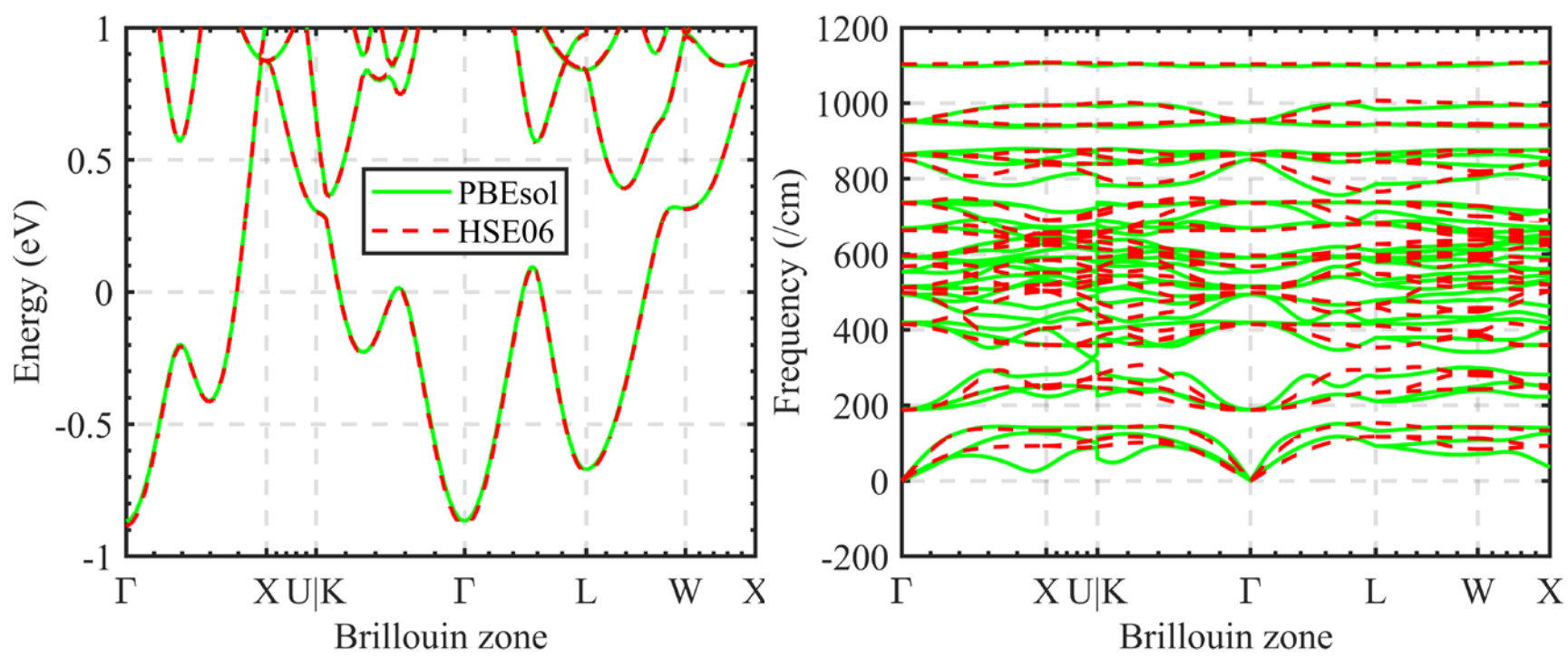

**Figure 5. The electronic band structure and phonon spectrum of $Rb(BN)_5$ at the PBEsol and HSE06 levels, respectively.**

It is noteworthy that even a slight pressure of 2 GPa can dramatically enhance the $T_c$ of $Rb(BN)_5$ from 113 K to 138 K at the PBEsol level, reaching a record comparable to that of copper oxides.[73] The physical mechanism underlying such a remarkable $T_c$ enhancement of 25 K under a modest pressure of only 2 GPa merits thorough investigation. It is well established that achieving high-temperature superconductivity necessitates large values of both the $\lambda$ and the $\omega_{\log}$. However, these two parameters typically exhibit a subtle trade-off relationship. As the hydrostatic pressure increases from 0 to 2 GPa, $\lambda$ decreases from 2.95 to 2.46 (a 16.7% reduction), whereas $\omega_{\log}$ surges from 274.9 K to 392.27 K—a substantial gain of 42.7%. Consequently, the positive boost imparted to $T_c$ by the stiffening of $\omega_{\log}$ far outweighs the adverse suppression caused by the diminished $\lambda$. This pronounced low-pressure enhancement of $\omega_{\log}$ is intimately tied to the non-linear lattice compressibility. In the low-pressure regime, the crystal exhibits higher compressibility, wherein a given pressure increment $\Delta P$ induces a significantly larger volumetric reduction than in the stiffer high-pressure regime. This volumetric response naturally explains why the slope of $\omega_{\log}$ with respect to pressure $d\omega_{\log}/dP$ progressively diminishes with increasing pressure (Figure 4(b)). Although the finding that the PBE functional predicts a maximum $T_c$ of up to 155 K at 16 GPa is also exciting, benchmarking against the pressure-volume equation of state reveals that the PBEsol calculations more closely align with the $r^2$SCAN+rVV10 (Figure 4(b)),[74] implying greater reliability of its results. Furthermore, the electronic structure of $Rb(BN)_5$ at 0 GPa is consistent with that obtained using the hybrid functional HSE06 (Figure 5). Additionally, the phonon spectrum derived from the supercell frozen-phonon method also demonstrates dynamical stability (Figure 5), indicating that higher theoretical levels do not undermine the conclusions drawn from the current PBEsol calculations. It is noteworthy that the consistent electronic band structures and similar phonon spectra presented in Figure 5 reveal that more advanced methods may not significantly affect the estimation of the EPC strength $\lambda$ (Figure S5). Moreover, the inner-shell electron effects under extreme pressure are likely the cause of discrepancies in electronic structure estimations for $CeH_9$.[75]

## 3. **Conclusions**

In summary, this work systematically investigates XC effects on covalent-cage BCS superconductor predictions, highlighting the marked sensitivity of functional choice and its impact on superconducting potential assessment. PBE's intrinsic over-response to electron density gradients systematically expands lattices, underestimating force constants, softening phonons, and overestimating electron-phonon coupling—critically compromising dynamic stability evaluation. This obscures high-$T_c$ potential for $K(BN)_5$ of 102 K and $Rb(BN)_5$ of 113 K at ambient pressure, with $Rb((BN)_5$ reaching 138 K at merely 2 GPa. XC-induced stabilization operates through dual pathways: (i) lattice-scale phonon hardening from stronger exchange interactions, and (ii) electronic-scale band widening reducing Van Hove singularities and $\sigma^*$ antibonding state accumulation at the Fermi level, suppressing Peierls distortion. This renders short-range covalent bonds exceptionally XC-sensitive, distinct from conventional metallic-bond superconductors. Pressure-dependent analysis reveals that high pressure attenuates phonon property discrepancies between functionals, whereas electronic property divergence persists. Phonon "robustness" stems from pressure-reduced XC energy contributions and homogenized electron density; fundamental functional differences maintain systematic biases, rendering Fermi-level bands exceptionally sensitive to XC uncertainty. Our findings highlight PBEsol as a suitable functional for providing accurate structural baselines in covalent-cage superconductors, providing critical benchmarks for high-throughput screening and refining computational materials design paradigms.

## Declaration of Competing Interest

The authors declare that they have no known competing financial interests or personal relationships that could have appeared to influence the work reported in this paper.

## Acknowledgements

The work described in this paper was supported by grants from the Research Grants Council of the Hong Kong SAR (No. 11317122). The authors also thank for the support of Open Source Supercomputing Center of Scientific-Advanced-Intelligence (SAI) for providing the high-performance simulation.

## Supplementary materials

Supplementary materials to this article can be found online.